\documentclass[preprint,preprintnumbers,amsmath,amssymb]{revtex4}
\usepackage{graphicx}
\usepackage{dcolumn}
\usepackage{bm}
\usepackage[dvips]{color}

\newcommand{\beq}{\begin{equation}}
\newcommand{\eeq}{\end{equation}}
\newcommand{\beqn}{\begin{eqnarray}}
\newcommand{\eeqn}{\end{eqnarray}}
\newcommand{\dda}[2]{#1^{\dag}_{#2}}
\newcommand{\da}[2]{#1_{#2}}

\begin{document}

\title{Persistent spin current in anisotropic spin ring}

\author{Yin Cheng and You-Quan Li}
\affiliation{Zhejiang Institute of Modern
Physics, Zhejiang University, Hangzhou 310027, P.R. China}
\author{Bin Chen}
\affiliation{Department of Physics, Hangzhou Teacher's College, Hangzhou 310036, P.R. China}

\begin{abstract}

The persistent spin current in anisotropic spin ring penetrated
by a SU(2) flux is studied by the Schwinger-boson mean field approach.
The anisotropy in spin coupling can facilitate the persistent spin current.
Ground-state energy and excitation energy gap are also studied.
The peak of spin current occurs at the maximum value of the
ground-state energy.

\end{abstract}


\received{\today}

\maketitle

\section{Introduction}

The quantum coherence plays a central role in mesoscopic physics and
the persistent current on mesoscopic rings
threaded by a magnetic flux is a particular sensitive probe of
such coherence. Thus there has been much study on persistent electrical current in
mesoscopic ring both experimentally~\cite{Levy,Chand,Mail}
and theoretically~\cite{Butt,Amb,LiMa,Koskinen}.
Owning to recent interests in the spin based electronics~\cite{Awsch},
the study on spin current
becomes a remarkable topic~\cite{Zhang,Niu0403,Halperin,JinLiZhang,Spletts}.
The persistent spin current in the ferromagnetic Heisenberg ring  was shown to occur
in the presence of crown-shaped magnetic field~\cite{Schutz1}.
It can also be driven by inhomogeneous
electric fields~\cite{Cao} due to the Aharonov-Casher effect~\cite{Aharonov}.
On the basis of spin-wave approach, spin current in antiferromagnetic
Heisenberg ring in inhomogeneous magnetic field
has been investigated very recently~\cite{Schutz2}.
It is well-known that the electrical persistent current is an
topological current produced by the magnetic flux, a "U(1) flux".
However, as far as we were aware, there is not a thorough discussion
about persistent spin current in the anisotropic Heisenberg model produced
by a SU(2) flux.

In this paper,  we study the anisotropy Heisenberg rings (XXZ
model) penetrated by a SU(2) flux.
In Sec.\ref{sec:Schwinger}, we apply the Schwinger-boson approach to the model.
In Sec.\ref{sec:GSE} we calculate the excitation spectrum and obtain
the ground-state energy and energy gap.
In Sec.\ref{sec:PC}, we evaluate the persistent spin current and discuss
the effects caused by the anisotropy in the model.
In Sec.\ref{sec:summary}, we give a summary of our main results.

\section{Schwinger-boson approach}\label{sec:Schwinger}

We consider a spin ring with anisotropy penetrated by the SU(2) flux:
\beqn\label{eq:SpinModel}
H=\sum^N_j\big[\,\frac{1}{2}(e^{i\phi/N}s^{+}_js^{-}_{j+1}
  +e^{-i\phi/N}s^{-}_js^{+}_{j+1})
\Delta s^z_js^z_{j+1}\big]+{\it h.c.},
\eeqn
where $s^{\pm}_j$ denote for spin flipping operarors and $s^z_j$ for
the $z$ component of spin operator;
$\phi=\Phi/\Phi_0$ with  $\Phi$ and $\Phi_0$ ($=hc/e$) the SU(2)
flux and the flux quanta respectively; $N$ the lattice number.
The anisotropy is characterized by the parameter $\Delta$. As is known that
$\Delta=-1$ corresponds to the ferromagnetic case, $\Delta=1$ is the
anti-ferromagnetic regime and $-1<\Delta<1$ is the transition regime from ferromagnetic
to anti-ferromagnetic.

In terms of Schwinger boson operators $a_j$ and $b_j$ which satisfy the
Bose commutation relations
$[a_i, a^{\dag}_j]=\delta_{ij}$ and $[a_i,
b_j]=0$, the spin operators are given by
\beq
s^{+}_j=a^{\dag}_jb_j,\quad\,s^{-}_j=a_jb^{\dag}_j,\quad\,s^z_j
 =\frac{1}{2}(a^{\dag}_ja_j-b^{\dag}_jb_j)
\eeq
with a local constraint at every site \emph{j} given by
$a^{\dag}_ja_j+b^{\dag}_jb_j=2S$ which means only $2S$  of the two bosons  can
occupy each site.

Since the lattice is a bipartite lattice, we can make a unitary transformation
$a_{j+1}\rightarrow -a_{j+1}\,,\,\,b_{j+1}\rightarrow b_{j+1}$ at each site of one
sublattice. This brings about
$s^{\pm}_{j+1}\rightarrow-s^{\pm}_{j+1}$
and $s^z_{j+1}\rightarrow s^z_{j+1}$.
The XXZ Hamiltonian (\ref{eq:SpinModel}) becomes the following form:
\begin{widetext}
\beqn
H&=&\frac{1}{2}\sum_j\Big[\,\Big(-e^{i\phi/N}\dda{a}{j}\da{b}{j}\da{a}{j+1}\dda{b}{j+1}
-e^{-i\phi/N}\da{a}{j}\dda{b}{j}\dda{a}{j+1}\da{b}{j+1}\Big),
\nonumber \\
&+&\frac{\Delta}{2}\Big(\dda{a}{j}\da{a}{j}\dda{a}{j+1}\da{a}{j+1}-\dda{a}{j}\da{a}{j}\dda{b}{j+1}\da{b}{j+1}-\dda{b}{j}\da{b}{j}\dda{a}{j+1}\da{a}{j+1}
+\dda{b}{j}\da{b}{j}\dda{b}{j+1}\da{b}{j+1}\Big)\Big]+{\it h.c.}
\eeqn
\end{widetext}
To fulfil the constraint, we need to introduce
a Lagrangian-multiplier field $\lambda_i$.
Then a generalized Hamiltonian is obtained
\begin{widetext}
\beqn
H=-\sum_j\Bigg\{\!\Bigg[\frac{1-\Delta}{4}\,{\cal
A}^{\dag}_{j,j+1}{\cal A}_{j,j+1}+\frac{1+\Delta}{4}\,{\cal
B}^{\dag}_{j,j+1}\da{{\cal B}}{j,j+1}\Bigg]\!\!+{\it h.c.}\Bigg\}\!\!
\nonumber \\
+2\sum_j\lambda_i\Big(\dda{a}{j}\da{a}{j}+\dda{b}{j}\da{b}{j}-2S\Big)+2NS^2+(1-\Delta)NS.
\eeqn
\end{widetext}
where
\begin{eqnarray*}
&&{\cal A}^{\dag}_{j,j+1}=e^{-i\phi/2N}\da{a}{j}\dda{a}{j+1}+e^{i\phi/2N}\da{b}{j}\dda{b}{j+1},
\nonumber \\
&&{\cal B}^{\dag}_{j,j+1}=e^{i\phi/2N}\dda{a}{j}\dda{b}{j+1}+e^{-i\phi/2N}\dda{b}{j}\dda{a}{j+1}.
\end{eqnarray*}
The isotropy limits $\Delta=\pm 1$ ({\it i.e.}, ferro- and antiferromagnetic cases)
without flux were considered in Ref.~\cite{Charles}.
At the mean-field level, we take the average value of
the multiplier field $\langle \lambda_i\rangle=\lambda$ and make
the bond operators
$\langle \frac{1-\Delta}{4}\dda{{\cal A}}{j,j+1}\rangle=A^{\ast}$ and
$\langle \frac{1+\Delta}{4}\dda{{\cal B}}{j,j+1}\rangle=B^{\ast}$
uniform and static. We hence obtain the mean-field Hamiltonian:
\begin{widetext}
\beqn
H_{\textmd{MF}}&=&-\sum_j\bigg\{\Big[\big(A^{\ast}\cdot{\cal A}_{j,j+1}
+A\cdot{\cal A}^{\dag}_{j,j+1}\big)+\big(B^{\ast}\cdot{\cal B}_{j,j+1}
+B\cdot{\cal B}^{\dag}_{j,j+1}\big)\Big]+{\it h.c.}\bigg\},
\nonumber \\
&+&2\lambda\sum_j(\dda{a}{j}\da{a}{j}+\dda{b}{j}\da{b}{j}-2S)
+\frac{8}{1-\Delta}A^{\ast}AN+\frac{8}{1+\Delta}B^{\ast}B+2NS^2+(1-\Delta)NS.
\eeqn
\end{widetext}
By making use of a Fourier transform
$\da{a}{j}=\Sigma_{\mathbf k} \,a_{\mathbf k} \exp(i\mathbf k r_j)$
where the summation $\mathbf k$ runs over the first Brillouin zone,
a four-component spinor
\beq \dda{\Psi}{\mathbf k}
 =(\dda{a}{\mathbf k}\,,\,\da{a}{\mathbf k}\,,\,\dda{b}{-\mathbf k}\,,\,\da{b}{-\mathbf k}),
\eeq
can be introduced. We can write the mean-field Hamiltonian into a compact form in the momentum space:
\beqn
H_{\textmd{MF}}=\sum_{\mathbf k}\bigg\{\dda{\Psi}{\mathbf k}\Big[\lambda-2\cos\Big(\mathbf k+\frac{\phi}{2N}\Big){\cal M}\Big]\Psi_{\mathbf k}
+2A\cos\Big(\mathbf k+\frac{\phi}{2N}\Big)\bigg\}+\varepsilon_0,
\label{eq:MFHamiton}
\eeqn
where
\begin{eqnarray*}
\varepsilon_0=\frac{8}{1-\Delta}A^{\ast}AN+\frac{8}{1+\Delta}B^{\ast}BN+(1-\Delta)NS
+2NS^2-2\lambda N(2S+1),
\end{eqnarray*}
and
\begin{eqnarray*}
{\cal M}=
\left(%
\begin{array}{cccc}
  A^{\ast} & 0 & 0& B\\[1mm]
  0 & A & B^{\ast} & 0 \\[1mm]
  0 & B & A^{\ast} & 0 \\[1mm]
  B^{\ast} & 0 & 0 & A \\[1mm]
\end{array}%
\right).
\end{eqnarray*}

The Hermitian property of the Hamiltonian (\ref{eq:MFHamiton}) enables us easily to
obtain that $A=A^{\ast}$.
Using a Bogoliubov transformation given by the following transformation matrix ${\cal T}$:
\beq
{\cal T}=
\left(%
\begin{array}{cccc}
  u \;\;\; & 0\;\;\; & 0\;\;\; & v \\[1mm]
  0\;\;\; & u\;\; & v\;\;\; & 0 \\[1mm]
  0\;\;\; & v\;\;\; & u\;\;\; & 0 \\[1mm]
  v\;\;\; & 0\;\;\; & 0\;\;\; & u \\[1mm]
\end{array}%
\right), \eeq
we transform the original Bose operators $\{\dda{a}{\mathbf
k}\,,\,\da{a}{\mathbf k}\,,\,\dda{b}{-\mathbf
k}\,,\,\da{b}{-\mathbf k}\}$ to a set of new Bose operators,
called ``quasi-particle" creation/anihilation operators,
$\{\dda{\alpha}{\mathbf k}\,,\,\da{\alpha}{\mathbf
k}\,,\,\dda{\beta}{\mathbf k}\,,\,\da{\beta}{\mathbf k}\}$
\beq {\cal T}
\left(%
\begin{array}{c}
  \dda{a}{\mathbf k} \\[0.5mm]
  \da{a}{\mathbf k} \\[0.5mm]
  \dda{b}{-\mathbf k} \\[0.5mm]
  \da{b}{-\mathbf k} \\[0.5mm]
\end{array}%
\right)
=
\left(%
\begin{array}{c}
  \dda{\alpha}{\mathbf k} \\[0.5mm]
  \da{\alpha}{\mathbf k} \\[0.5mm]
  \dda{\beta}{\mathbf k} \\[0.5mm]
  \da{\beta}{\mathbf k} \\[0.5mm]
\end{array}%
\right).
\eeq
Then the Hamiltonian is diagonalized,
\beqn
&&H_{\textmd{MF}}\!\!=\!\sum_{\mathbf k}\!\bigg[\omega_{\mathbf k,\phi}(\dda{\alpha}{\mathbf k}\da{\alpha}{\mathbf k}+\dda{\beta}{k}\da{\beta}{\mathbf k}+1)
+2A\cos\Big(\mathbf k+\frac{\phi}{2N}\Big)\!\bigg]+\varepsilon_0,
\label{eq:Hamilton}
\eeqn
where the quasi-particle spectrum is
\beq
\omega_{\mathbf k,\phi}\!=\!\sqrt{\!\Big[\lambda-2A\cos\big(\mathbf k+\frac{\phi}{2N}\big)\!\Big]^2\!\!\!-\Big|2B\cos\big(\mathbf k+\frac{\phi}{2N}\big)\!\Big|^2}.
\label{eq:specture}
\eeq
Thus the free energy is obtained
\beqn
f=\frac{F_{\textmd{MF}}}{2N}=\int^{\frac{\pi}{2}}_{-\frac{\pi}{2}}\frac{\textmd{d}\mathbf k}{2\pi}\bigg[\frac{2}{\beta}\ln\Big(2\sinh\frac{\beta
\omega_{\mathbf k,\phi}}{2}\Big)
+2A\cos\Big(\mathbf k+\frac{\phi}{2N}\Big)\bigg]+\frac{\varepsilon_0}{2N}.
\label{eq:freeenergy}
\eeqn
where $\beta=1/k_B T$ with $k_B$ the Boltzmann constant and $T$ the temperature.
The mean-field self-consistent equations are obtained by minimizing the free energy,
{\it i.e.}, $\delta f/\delta A=0\,,\,\delta f/\delta B=0\,,\delta f/\delta \lambda=0$,
then the saddle-point equations are given by
\begin{widetext}
\beqn
&&A+\frac{1-\Delta}{4}\int^{\frac{\pi}{2}}_{-\frac{\pi}{2}}\frac{\textmd{d}\mathbf k}{2\pi}\bigg[1-\frac{\lambda-2A\cos\Big(\mathbf k+\frac{\phi}{2N}\Big)}{\omega_{\mathbf k,\phi}}\cdot\coth\Big(\frac{\beta \,\omega_{\mathbf k,\phi}}{2}\Big)\bigg]\cdot\cos\Big(\mathbf k+\frac{\phi}{2N}\Big)=0,
\nonumber  \\
&&1-\frac{1+\Delta}{2}\int^{\frac{\pi}{2}}_{-\frac{\pi}{2}}\frac{\textmd{d}\mathbf k}{2\pi}\bigg[\frac{\big|\cos(\mathbf k+\frac{\phi}{2N})\big|^2}{\omega_{\mathbf k,\phi}}\cdot\coth\big(\frac{\beta\,\omega_{\mathbf k,\phi}}{2}\big)\bigg]=0,
\nonumber \\
&&(2S+1)-\int^{\frac{\pi}{2}}_{-\frac{\pi}{2}}\frac{\textmd{d}\mathbf k}{2\pi}\bigg[\frac{\lambda-2A\cos\big(\mathbf k+\frac{\phi}{2N}\big)}{\omega_{\mathbf k,\phi}}\cdot\coth\Big(\frac{\beta\,\omega_{\mathbf k,\phi}}{2}\Big)\bigg]=0,
\label{eq:saddle}
\eeqn
\end{widetext}
which determine the parameters $A$, $B$  and the Lagrange multiplier (chemical potential)
$\lambda$ in Eqs.~(\ref{eq:specture})-(\ref{eq:freeenergy}).
These saddle point equations can be solved numerically.

\section{The ground-state energy and energy gap}\label{sec:GSE}

We consider $S=1$ at $T=0\,\textmd{K}$ with $N=50$ sites. At zero
temperature, the ground-state (GS) energy equals to the free
energy because there is no thermal fluctuation. When
$T\!\rightarrow0^+$, $\beta$ tends to $+\infty$ which means
$e^{-\beta\omega_{\mathbf k,\phi}/2}\rightarrow0^+$ if
$\omega_{\mathbf k,\phi}$ is finite. Thus in the first term of the
integral in the free energy equation (\ref{eq:freeenergy}),
$\ln(2\sinh\frac{\beta}{2}\omega_{\mathbf k,\phi})$ reduces to
$\omega_{\mathbf k,\phi}$ and the free energy ({\i.e.},
ground-state energy per site at $T=0$) becomes
\beq
E_{\textmd{GS}}(\phi)=\int^{\frac{\pi}{2}}_{-\frac{\pi}{2}}\frac{\textmd{d}\mathbf
k}{2\pi}\bigg[\omega_{\mathbf k,\phi}+2A \cos\Big(\mathbf
k+\frac{\phi}{2N}\Big) \bigg]+\frac{\varepsilon_o}{N},
\label{eq:groundenergy}
\eeq
We plot the ground-state energy per
site versus $\theta$ (here we set $\phi/N=\theta$) for
$\Delta=0.9$ in Fig. \ref{fig:GS}. Apparently, the ground-state
energy has a sharp cusp at $\theta/2\pi=\pm 0.5$ whose magnitude
is $-0.9626$, which is higher than average value of the energy
gap. The height of cusp peak raises while $\Delta$ decreases,
which is shown in the inset of the Fig.~\ref{fig:GS}. In other
words, more anisotropy of spin chain results in sharper cusp in
the curve.

\begin{figure}
\setlength{\unitlength}{0.8mm}
\includegraphics[width=7.8cm]{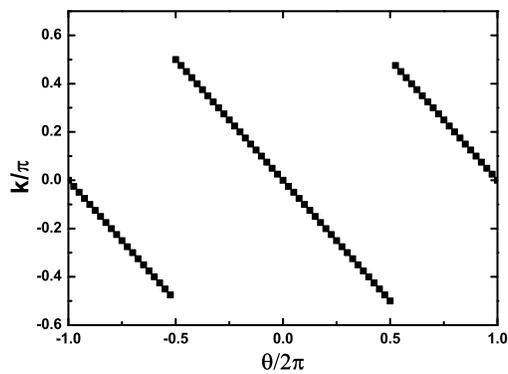}
\caption{\label{fig:ThetaK}
Momentum versus flux for $\Delta=0.9$.}
\end{figure}

\begin{figure}
\setlength{\unitlength}{0.8mm}
\includegraphics[width=7.8cm]{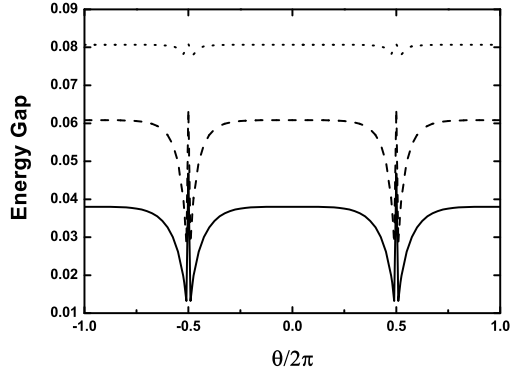}
\caption{\label{fig:EG}
The excitation energy gap versus flux for $\Delta=0.1$ (solid), $\Delta=0.5$ (dash)
and $\Delta=0.9$ (dot).}
\end{figure}

We calculate the energy gap (EG) of the first excited state over the ground state.
Unlike the usual case, the energy gap is not at $\mathbf k =0$
due to the presence of the external SU(2) flux.
This implies that the magnitude of the gap changes with the flux accordingly,
which is shown in Fig. {\ref{fig:ThetaK}}.
Clearly, $\mathbf k$ undergoes a jump at $\theta\!/2\pi=\pm 0.5$.
The Fig. \ref{fig:EG} shows the excitation energy gap versus the magnetic flux.
The gap descents as the flux goes to
$\theta\!/2\pi=\pm 0.5$, but rebounds when it is very close to $\theta\!/2\pi=\pm 0.5$
and reaches the maximum at $\theta\!/2\pi=\pm 0.5$.

\section{Persistent spin current}\label{sec:PC}

\begin{figure}
\setlength{\unitlength}{0.8cm}
\includegraphics[width=7.8cm]{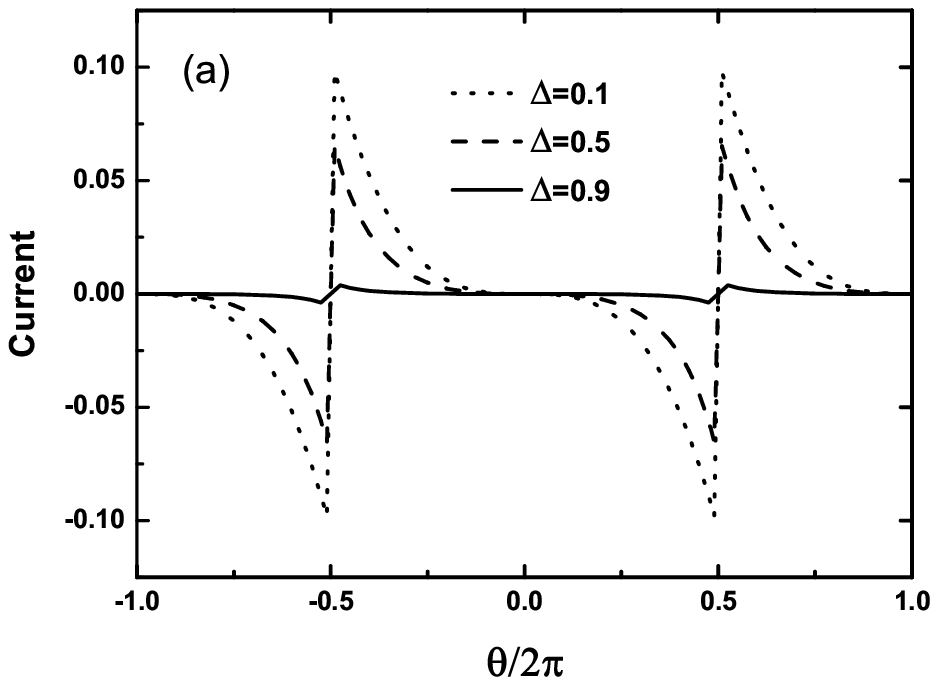}
\includegraphics[width=7.8cm]{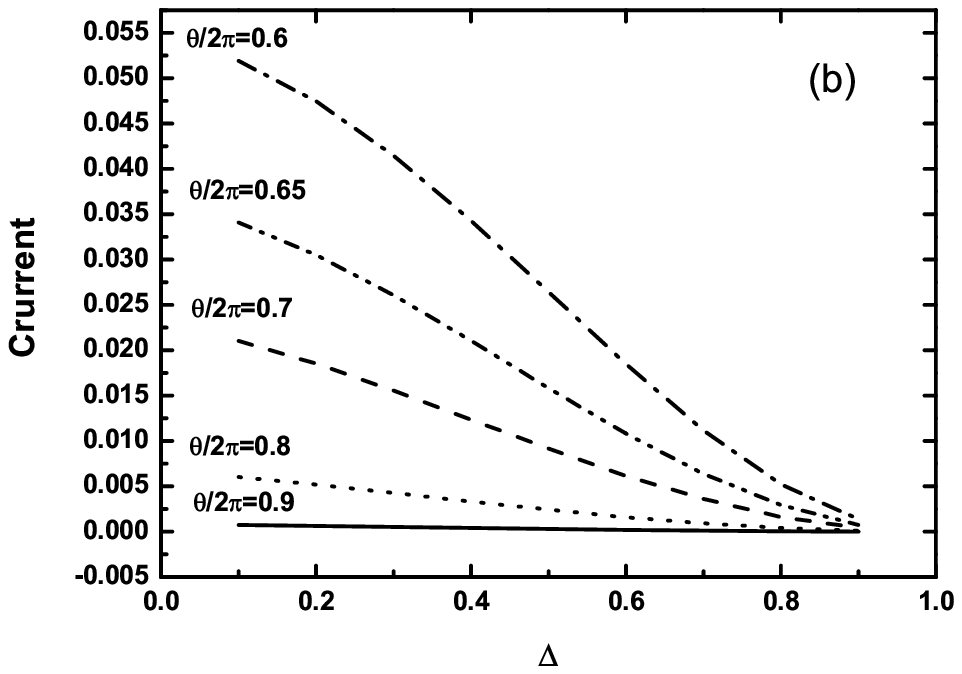}
\caption{\label{fig:persistent}
(a) Persistent spin current versus flux for various anisotropy parameter $\Delta$.
The dash-, solid-, and dot-line corresponds to $\Delta=0.1,\;0.5$ and $0.9$ respectively.
(b) The current versus $\Delta$ illustrates the effects caused by anisotropy parameter.}
\end{figure}

\begin{figure}
\setlength{\unitlength}{0.8mm}
\includegraphics[width=7.6cm]{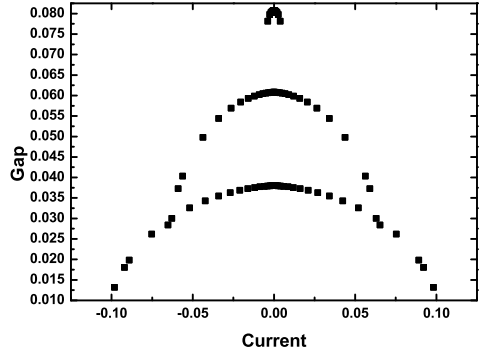}
\caption{\label{fig:PSGap}
Energy gap versus persistent spin current for different anisotropy parameters $\Delta$.
Hollow circle($\circ$), cross($\times$) and solid circle($\bullet$) corresponds to
$\Delta=0.1,\;0.5$ and $0.9$ respectively.}
\end{figure}

We have previously obtained the ground-state energy and excitation
spectrum whose values are determined by Eqs.(\ref{eq:saddle}). Now
we are in the position to evaluate the persistent current at zero
temperature ($T=0\,\textmd{K}$) which is defined by
\beq
I(\phi\,,T)=-\frac{\partial E_{\textmd{GS}}(\phi)}{\partial\phi},
\label{eq:SpCurrent}
\eeq
where $E_{\textmd{GS}}(\phi)$ is the
ground-state energy Eq.~(\ref{eq:groundenergy}). The charge
current $I_c$ is null in our model for the particle is fixed on
each site. However, the pure persistent spin current in the ring
can be deduced by the SU(2) flux. We plot numerical calculation of
Eq. (\ref{eq:SpCurrent}) in Fig.~\ref{fig:persistent} (a). The
jumps in the spin current curve occur at $\theta/2\pi=\pm 0.5$
for various $\Delta$. We found that
the anisotropy ($\Delta$) will enhance the persistent spin
current, which is shown in (b) of Fig.\ref{fig:persistent}. As a
result, the spin current in $XY$ limit is larger than  in
Heisenberg limit. Fig.~\ref{fig:PSGap} exhibits that the energy
gap decreases while persistent current increases. The larger the
energy gap is, the more spin flippings that contributes to spin
current are prevented.

\section{Summary and discussion}\label{sec:summary}

\begin{figure}
\setlength{\unitlength}{0.8mm}
\includegraphics[width=7.8cm]{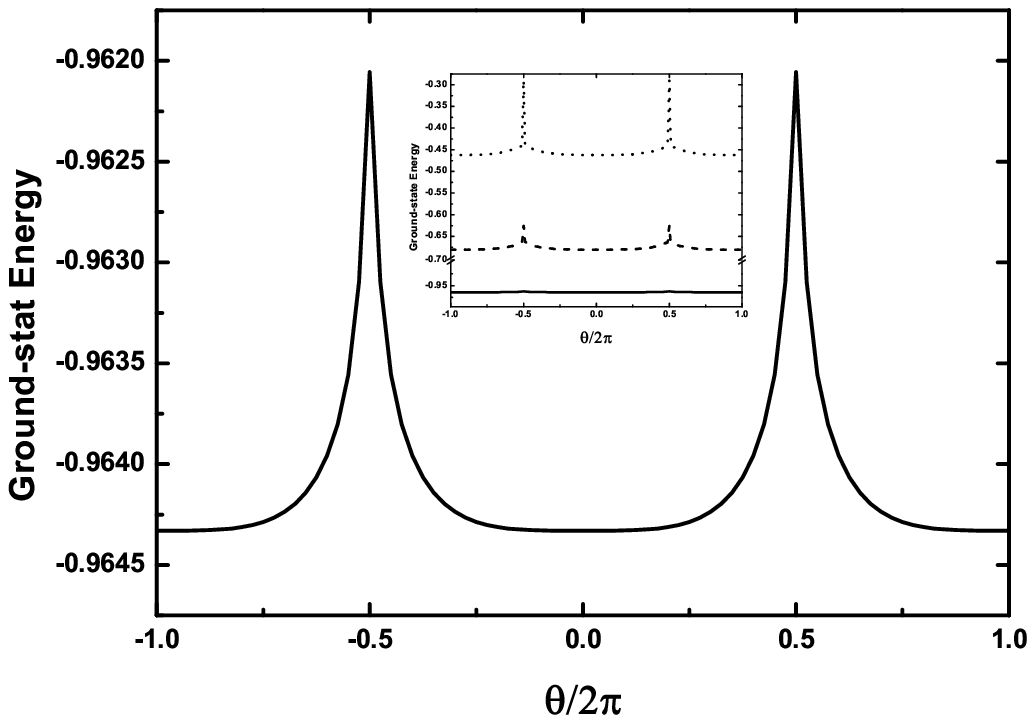}
\caption{\label{fig:GS} The ground-state energy versus SU(2) flux $\theta$
($\theta=\phi/N$) for $\Delta=0.9$.
The insert shows ground-state energy for various anisotropy
$\Delta=0.1$ (dot), $\Delta=0.5$ (dash) and $\Delta=0.9$ (solid).}
\end{figure}

Using Schwinger-bonson mean field approach, we have investigated the
property of ground-state and energy gap for the anisotropy spin ring
penetrated by SU(2) flux.
In the curve of energy versus flux, there is a cusp
at $\theta/2\pi=\pm 0.5$ and the energy reaches a maximum.
Whereas, the excitation energy gap drops drastically nearby $\theta/2\pi=\pm0.5$
but rebound to maxima at $\theta/2\pi=\pm0.5$.
This implies that the energy curve of the first excited state has
a similar shape as Fig.~\ref{fig:GS} and is tangent to the curve of the ground state
near the point $\theta=\pm 0.5$.
We calculated the pure persistent spin current and found that
the flux dependence of persistent spin currents are facilitated by the anisotropy parameter
$\Delta$ which promotes the persistent spin current.
The peak of spin current appears at the minimal value of excitation gap.

The work is supported by NSFC grant No.10225419.

\newpage

\end{document}